\documentclass[a4paper,11pt]{article}
\pdfoutput=1 

\usepackage{epsfig}
\usepackage{graphicx}                  
\usepackage{ae}
\usepackage{amsmath}
\usepackage{amssymb}
\usepackage{graphics}
\usepackage{dcolumn}
\usepackage{bm}
\usepackage{lineno}
\usepackage{caption}
\usepackage{ragged2e}

\title{\bf Observation on the bias current variation of a single mask triple GEM chamber}
\date{}

\begin{document}
		\maketitle
	\flushbottom
\vspace*{-1cm}
\centering
{
\author{S. Chatterjee$^{1*}$,}
\author{A. Sen$^{1}$,}
\author{R. Paul$^{2}$,}
\author{S. Sahai$^{3}$,}
\author{S. Das$^{1}$}
\author{and S. Biswas$^{1}$}

}

\vspace*{0.5cm}
$^1${Department of Physics, Bose Institute, EN-80, Sector V, Kolkata, India}
$^2${Department of Physics, Savitribai Phule Pune University, Pune, India}
$^3${Amity Institute of Nuclear Science and Technology, Noida, India}

\let\thefootnote\relax\footnotetext{$^*$Corresponding author. 

\hspace*{0.4cm}E-mail: sayakchatterjee896@gmail.com }
\vspace*{0.5cm}
\centering{\bf Abstract}
\justify
	Gas Electron Multiplier (GEM) detector, one of the advanced members of the Micro Pattern Gas Detector (MPGD) group, is widely used in High Energy Physics (HEP) experiments. The high rate handling capability and spatial resolution make it a desired tracking detector for high rate HEP experiments. Investigation of the long-term stability is an essential criterion for any tracking device used in HEP experiments. To investigate the long-term stability of a Single Mask~(SM) triple GEM detector prototype, it is irradiated continuously using a $^{55}$Fe X-ray source of energy 5.9 keV. The chamber is operated with Ar/CO$_2$ gas mixture in continuous flow mode. The gain and energy resolution of the chamber are calculated from the 5.9 keV X-ray peak and studied as a function of time. The applied voltage, divider current and also the environmental parameters (ambient temperature, pressure and relative humidity) are recorded continuously. It is observed that at a fixed applied voltage, the divider current of the detector is changing with time and as a result, the gain of the detector also changes. A systematic investigation is carried out to understand the probable reasons behind the observed variation in divider current and also to find its possible remedies. The details of the experimental setup, methodology and results are discussed in this article.



\section{Introduction}
\label{sec:intro}

The Gas Electron Multiplier~(GEM) detector is widely used in High Energy Physics~(HEP) experiments as a tracking detector due to its high rate handling capability~($\sim$ 1 MHz/mm$^2$) and good position resolution~($\sim$~70~$\mu$m)~\cite{GEM,GEM_foil}. GEM chambers are successfully used in several experiments like COMPASS~\cite{compass} at CERN SPS, LHCb~\cite{lhcb}, TOTEM~\cite{totem}, CMS~\cite{cms} and ALICE~\cite{alice} at LHC, CERN. Several future experiments such as CBM~\cite{cbm} at FAIR, NA60+~\cite{na60+} at CERN SPS are also considering the GEM detectors as the tracking chambers due to the foreseen high interaction rates. 

Long-term stability is of the utmost importance for any tracking detectors used in HEP experiments. Therefore, an initiative is taken to investigate the long-term stability of a SM triple GEM chamber with a $^{55}$Fe X-ray source of characteristic energy 5.9 keV. The chamber is operated with Ar/CO$_2$ gas mixture in 70/30 volume ratio in continuous flow mode. However, while investigating the long-term performance of the chamber by studying the gain variation as a function of time, a sudden decrease in gain is observed. In this article, the variation of gain as a function of time and its probable reasons are reported for the SM triple GEM chamber prototype.

\section{Experimental setup}

Long-term stability study is carried out with a SM triple GEM chamber prototype of dimension 10~cm~$\times$~10~cm. The drift gap, transfer gaps and induction gap of the chamber are kept fixed at 3~mm, 2~mm and 2~mm respectively~\cite{s_chatterjee}. 
The schematic of the HV distribution for the GEM chamber is shown in Fig.~\ref{fig2}~(top)~\cite{s_chatterjee}. A low pass HV filter is used between the input HV line and the resistive chain. A sum-up board is used to collect the signal from the detector. The output signal from the readout plane is fed to a charge sensitive pre-amplifier having a gain of 2~mV/fC. The output of the pre-amplifier is put to a linear Fan In Fan Out~(FIFO) module. One output from the FIFO is fed to a Multi Channel Analyzer~(MCA) to store the X-ray spectra in the computer. 
The ambient temperature, pressure and relative humidity are monitored continuously using a data logger built in house~\cite{data_logger}. The divider current is monitored continuously using the GEneral COntrol~(GECO) software~\cite{GECO} from CAEN. 

\section{Results and discussion}

The gain of the chamber is calculated by fitting the 5.9 keV main peak of the X-ray spectra using a Gaussian distribution. The relation between the MCA channel no. and the gain of the chamber can be expressed as;
\begin{equation}
	gain = \frac{((MCA~channel~no~\times~0.0014~+~0.1428~V)/2~mV)~fC}{No.~of~primary~electrons~\times~e~C}
\end{equation}
The details of the gain calculation are discussed in Ref.~\cite{s_chatterjee_1}. 
In order to nullify the effect of temperature~(T = t + 273 K) and pressure~(p in mbar) variations on the gain, it is normalized using the following relation~\cite{tp,pisa};

\begin{equation}
	normalized~gain~=~gain/A~exp(B~T/p)
\end{equation}   	
where A and B are the fit parameters obtained by studying the correlation of gain and T/p. After normalizing the gain to eliminate the temperature and pressure effects, the normalized gain is found to be decreased with time. While investigating the probable reason behind the decreasing trend of the normalized gain of the chamber, it is found that the divider current which is also monitored continuously using the GECO software~\cite{GECO}, decreases over time though the applied HV is kept constant. 
\begin{figure}[tbh!]
	\centering	
	\includegraphics[scale=0.35]{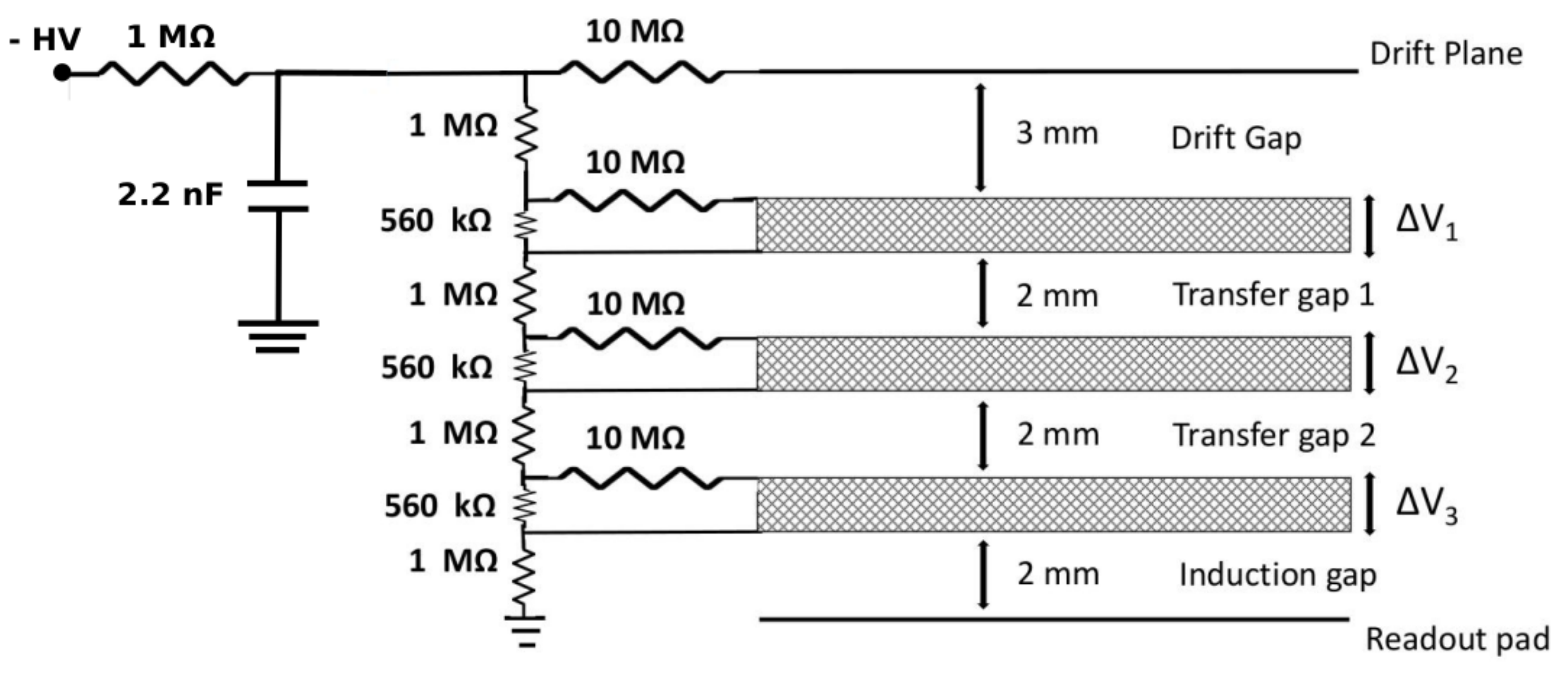}
	\includegraphics[scale=0.35]{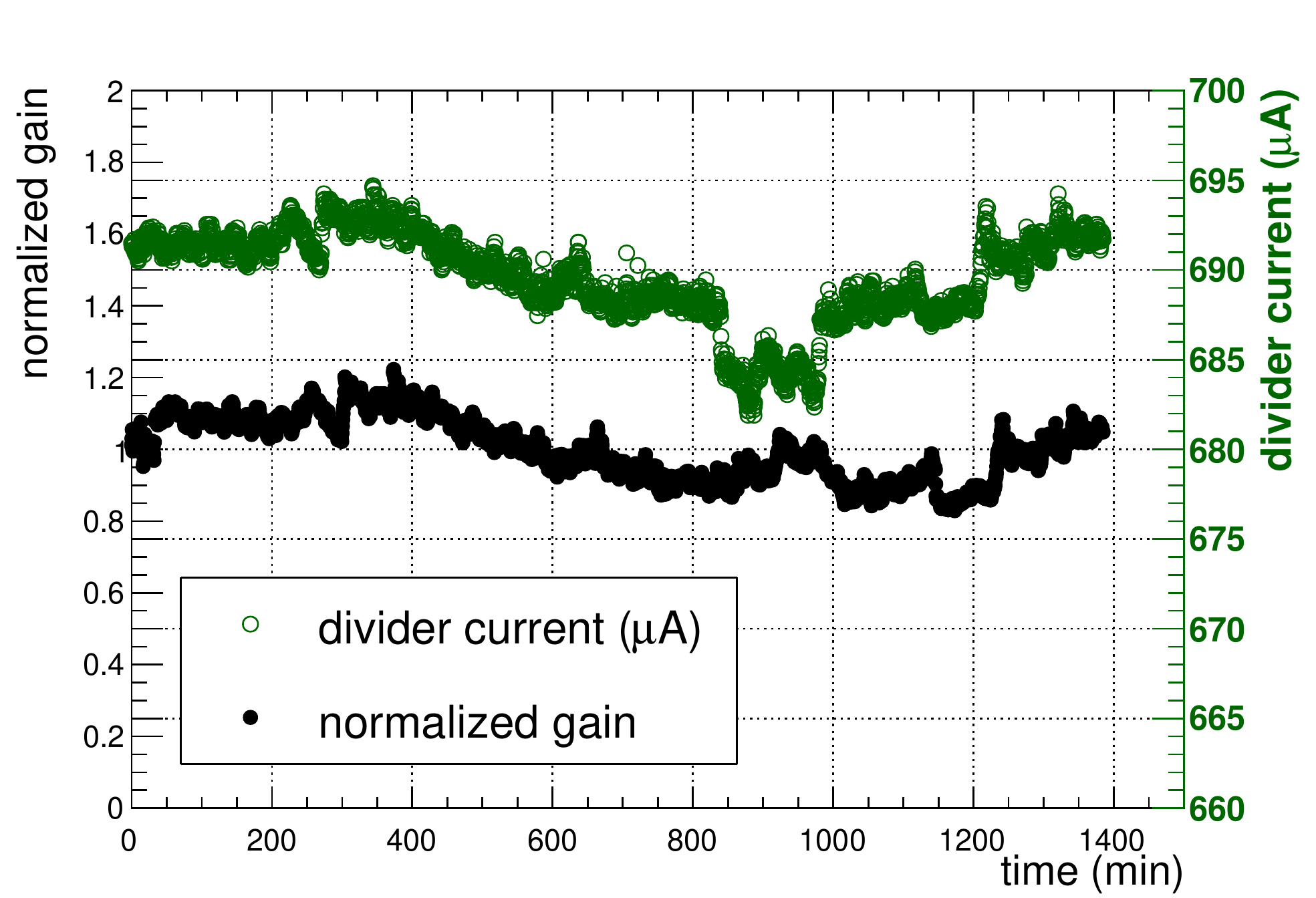}
	\caption{\label{fig2} Top: Schematic of the HV distribution across the GEM foils using the resistive chain network. Bottom: Variation of T/p normalized gain and divider current as a function of time. }
\end{figure} 
As the divider current decreases it reduces the potential difference~($\Delta$V) applied across the GEM foils and thus the gain of the chamber decreases. In Fig.~\ref{fig2}~(bottom) the variation of normalized gain and divider current is shown as a function of time. Within the time region of $\sim$~800 to 1000~minutes, a sudden drop in the divider current is observed and the probable reason for that is still not understood as this kind of behaviour is not observed repeatedly. However, from Fig.~\ref{fig2}~(bottom) a clear correlation as expected between the divider current and the normalized gain is observed except for the $\sim$~800-1000~minute time window because of the sudden drop in the divider current. The correlation between the divider current and normalized gain is shown in Fig.~\ref{fig3}~(top). The normalized gain is even further corrected for the divider current variation using the following relation;
\begin{equation}\label{eqn2}
	normalized~gain_{corrected}~=~normalized~gain/exp(p0~+~p1~\times~divider~current)
\end{equation}   
The variation of the normalized gain before and after correction for the divider current variation is shown in Fig.~\ref{fig3}~(bottom). 
\begin{figure}[tbh!]
	\centering
	\includegraphics[scale=0.35]{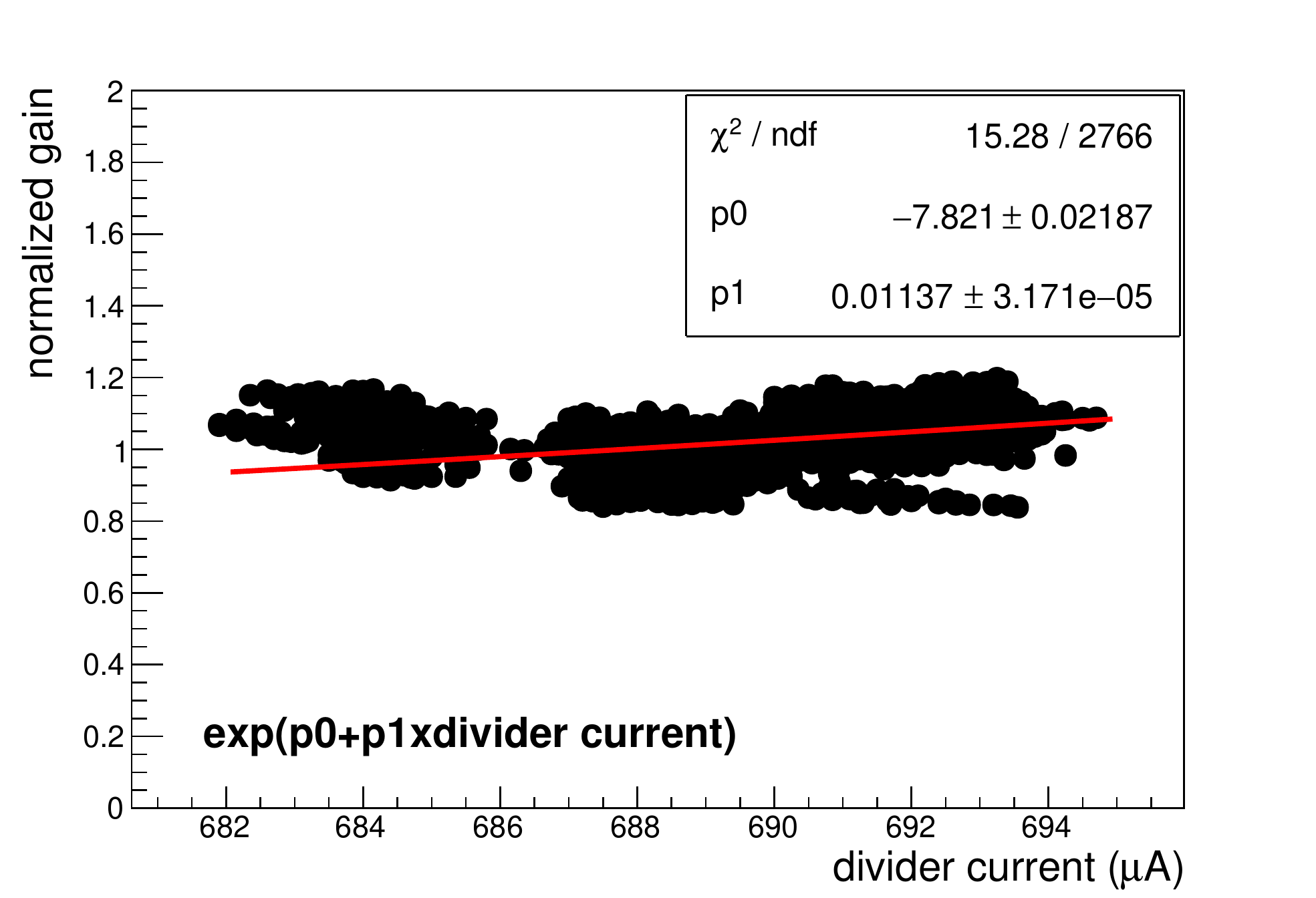}
	\includegraphics[scale=0.35]{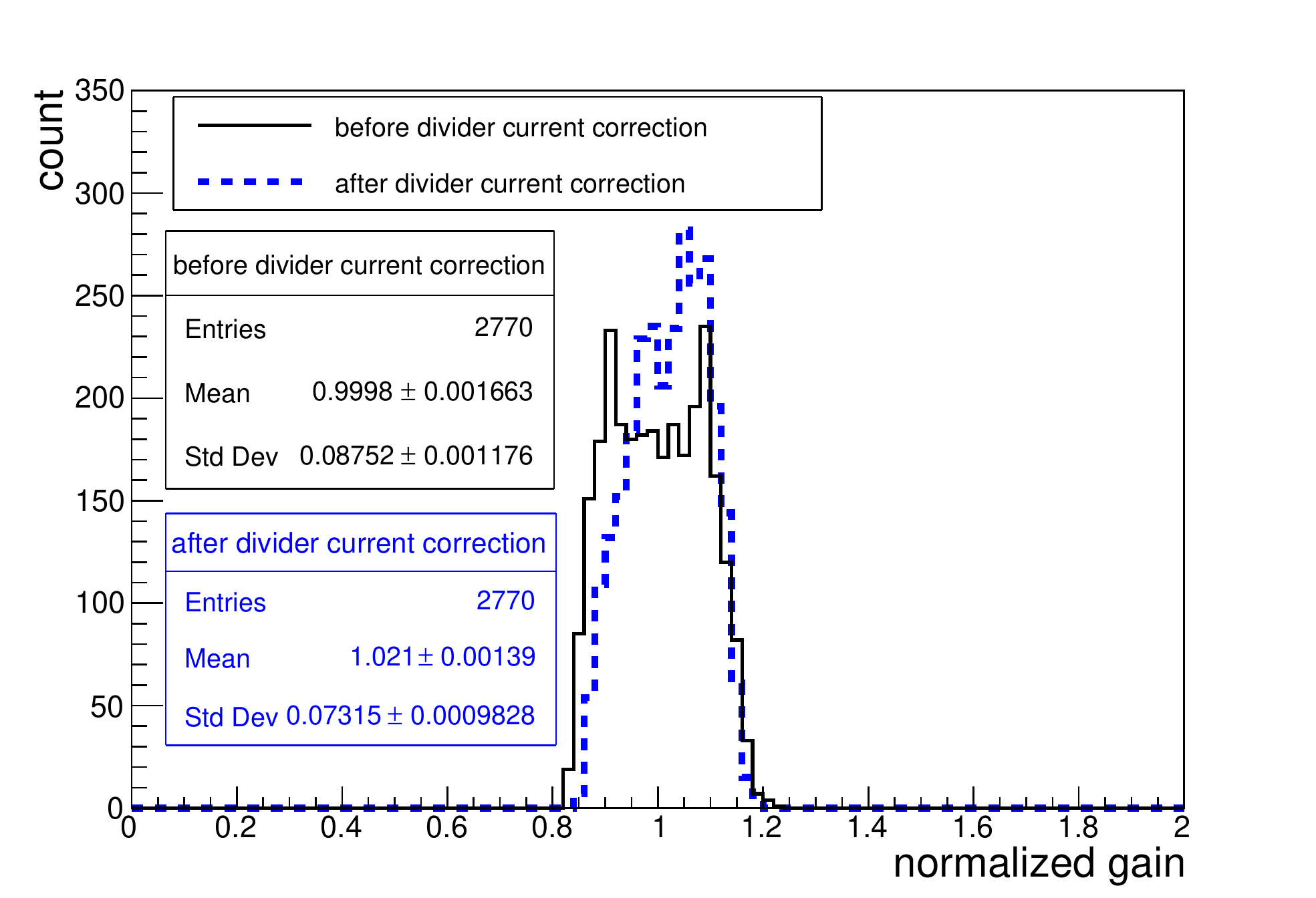}	
	\caption{\label{fig3}Top: Correlation between the divider current and T/p normalized gain. Bottom: Variation in normalized gain with~(blue line) and without correction~(black line) for the divider current variation.}
\end{figure} 
The uncorrelated portion~(between divider current of $\sim$~682-686~$\mu$A) in Fig.~\ref{fig3}~(top) is due to the sudden drop observed in the divider current during the time window of $\sim$~800-1000~minutes as shown in Fig.~\ref{fig2}~(bottom). The distribution of corrected normalized gain (using Eqn~\ref{eqn2}) shows a smaller standard deviation of $\sim$~7\% as compared to the standard deviation of $\sim$~9~\% obtained in the normalized gain when the correction for the divider current variation is not taken into account. 

Though the variation in the gain of the chamber over time is understood by studying the variation in the divider current at a fixed HV setting but the actual reason behind the observed variation in the divider current is still under investigation.

\section{Acknowledgment}

The authors would like to thank the RD51 collaboration for the support in building and initial testing of the chamber in the RD51 laboratory at CERN. S. Chatterjee would like to thank Mr. Pranjal Barik of Savitribai Phule Pune University, and Mr. Ayan Dandapat of IIT Ropar for their help in setting up the measurement setup. The authors would also like to thank Mr. Subrata Das of Bose Institute for helping in building the collimators used in this study. This work is partially supported by the research grant SR/MF/PS-01/2014-BI from DST, Govt. of India, and the research grant of
CBM-MuCh project from BI-IFCC, DST, Govt. of India. S. Biswas acknowledges the support of the DST-SERB Ramanujan Fellowship (D.O.No. SR/S2/RJN-02/2012).


\begin{thebibliography}{99}
	
	\bibitem{GEM} F. Sauli, Nucl. Instrum. Methods Phys. Res. A 386 (1997) 531
	
	\bibitem{GEM_foil}R. De Oliveira et al., United States Patent, Patent No.: US 8,597,490 B2, December 3, 2013
	
	\bibitem{compass}B. Ketzer et al.,  Nucl. Instrum. Methods Phys. Res. A, 535, 314 (2004)
	
	\bibitem{lhcb}M. Alfonsi et al., Nucl. Instrum. Methods Phys. Res. A, 518, 106 (2004)
	
	\bibitem{totem}S. Lami et al., Nuclear Physics B (Proc. Suppl.) 172, 231 (2007)
	
	\bibitem{cms}G. Mocellin, on behalf of the CMS Muon Group, J. Phys.: Conf. Ser. 1390, 012116 (2019)
	
	\bibitem{alice}ALICE TPC collaboration et al., J. Instrum. 16 P03022 (2021)
	
	\bibitem{cbm}S. Chattopadhyay et al., Technical Design Report for the CBM : Muon
	Chambers (MuCh); GSI, 2015 
	
	\bibitem{na60+}A. D. Falco for the NA60+ Collaboration, EPJ Web of Conferences 259, 09003 (2022)
	
	\bibitem{s_chatterjee}S. Chatterjee et al., Nucl. Instrum. Methods Phys. Res. A 1014 (2021) 165749
	
	\bibitem{data_logger}S. Sahu et al., J. Instrum 12 (2017)
	
	\bibitem{GECO}https://www.caen.it/products/geco2020/
	
	\bibitem{s_chatterjee_1}S. Chatterjee et al., J. Instrum. 15 (2020) T09011
	
	\bibitem{tp}M. C. Altunbas et al., Nucl. Instrum. Methods Phys. Res. A 515 (2003) 249
	
	\bibitem{pisa}S. Chatterjee et al., Nucl. Instrum. Methods Phys. Res. A 1046 (2023) 167747
	
\end{thebibliography}
\end{document}